\documentclass[fleqn,usenatbib]{mnras}

\usepackage{newtxtext,newtxmath}
\usepackage[T1]{fontenc}
\usepackage{ae,aecompl}

\usepackage{graphicx}	% Including figure files
\usepackage{amsmath}	% Advanced maths commands
\usepackage{amssymb}	% Extra maths symbols
\usepackage{ulem}
\newcommand{\angstrom}{\mbox{\normalfont\AA}}
\usepackage{titlesec}
\title[Detecting the Cosmic Web]{Detecting the Cosmic Web: Ly$\alpha$ Emission from Simulated Filaments at z=3}
\author[L. M. Elias et al.]
{\parbox{20cm}{
Lydia M. Elias$^{1}$\thanks{E-mail: lydia.elias@email.ucr.edu  (LME)},
Shy Genel$^{2,3}$,
Amiel Sternberg$^{2,4,5}$,
Julien Devriendt$^{6}$,
Adrianne Slyz$^{6}$,\\
Eli Visbal$^{7,8}$,
Nicolas Bouch\'e$^{9,10}$
}\vspace{0.3cm}\\
$^{1}$Department of Physics and Astronomy, University of California Riverside, 900 University Ave., Riverside, 92521\\
$^{2}$Center of Computational Astrophysics, 162 5th Ave., New York, NY, 10010\\
$^{3}$Columbia Astrophysics Laboratory, Columbia University, 550 West 120th Street, New York, NY 10027\\
$^{4}$School of Physics and Astronomy, Tel Aviv University, Ramat Aviv 69978, Israel\\
$^{5}$Max-Planck-Institut f\"{u}r extraterrestrische Physik (MPE), Gissenbachstr., D-85748, Garching, FRG\\
$^{6}$Department of Physics, University of Oxford, Keble Road, Oxford OX1 3RH, UK\\
$^{7}$ Department of Physics and Astronomy, University of Toledo, 2801 Bancroft Street, Toledo, OH, 43606\\
$^{8}$ Ritter Astrophysical Research Center, University of Toledo, 2801 Bancroft Street, Toledo, OH, 43606\\
$^{9}$ IRAP (Institut de Recherche en Astrophisique et Planetologie), Universit\'e de Toulouse, CNRS, UPS, F-31400 Toulouse, France\\ 
$^{10}$ Univ de Lyon, Univ Lyon1, Ens de Lyon, CNRS, Centre de Recherche Astrophysique de Lyon UMR5574, F-69230 Saint-Genis-Laval, France
}

\date{Accepted XXX. Received YYY; in original form ZZZ}

\pubyear{2020}

\def\Msun{\hbox{$\rm\thinspace M_{\odot}$}}
\def\AREPO{{\small AREPO}}

\begin{document}
\label{firstpage}
\pagerange{\pageref{firstpage}--\pageref{lastpage}}
\maketitle

\begin{abstract}
The standard cosmological model ($\Lambda$CDM) predicts the existence of the cosmic web: a distribution of matter into sheets and filaments connecting massive halos. However, observational evidence has been elusive due to the low surface brightness levels of the filaments. Recent deep MUSE/VLT data and upcoming observations offer a promising avenue for Ly$\alpha$ detection, motivating the development of modern theoretical predictions. We use hydrodynamical cosmological simulations run with the \AREPO{ }code to investigate the potential detectability of large-scale filaments, excluding contributions from the halos embedded in them. We focus on filaments connecting massive ($M_{\rm 200c}\sim(1-3)\cdot10^{12}\Msun$) halos at $z=3$, and compare different simulation resolutions, feedback levels, and mock image pixel sizes. We find increasing simulation resolution does not substantially improve detectability notwithstanding the intrinsic enhancement of internal filament structure. By contrast, for a MUSE integration of 31 hours, including feedback increases the detectable area by a factor of $\simeq 5.5$ on average compared with simulations without feedback, implying that even the non-bound components of the filaments have substantial sensitivity to feedback. Degrading the image resolution from the native MUSE scale of $(0.2")^2$ per pixel to $(5.3")^2$ apertures has the strongest effect, increasing the detectable area by a median factor of $\simeq$200 and is most effective when the size of the pixel roughly matches the width of the filament. Finally, we find the majority of Ly$\alpha$ emission is due to electron impact collisional excitations, as opposed to radiative recombination.
\end{abstract}

\begin{keywords}
large-scale structure of the Universe -- intergalactic medium -- diffuse radiation --methods: numerical
\end{keywords}

%%%%%%%%%%%%%%%%%%%%%%%%%%%%%%%%%%%%%%%%%%%%%%%%%%

%%%%%%%%%%%%%%%%% BODY OF PAPER %%%%%%%%%%%%%%%%%%

\section{Introduction}
 Cosmological simulations have long suggested the presence of diffuse filaments of both dark and baryonic matter between galaxies \citep{Peebles1975,Klypin1983,Haider2016,Mandelker2019}. However, it is only recently that the contribution of the cosmic web to galactic evolution has been fully appreciated. Specifically, at z$\geq$2, the primary mode of gas accretion for massive galaxies may be through streams of cold ($\sim10^4$ K) gas that feed the dark matter halo from cosmic web filaments \citep{Furlanetto2003,Katz2003,Keres2005,Dekel2006,Ocvirk2008,Dekel2009,vandeVoort2011,Nelson2016}.
%AMIEL The above statement needs to be qualified with more references added. The recent simulations imply a messy zone, not direct cold flow feeding of the disks. Shy, comments?

The low density of the filaments prohibits star formation, making observation extremely challenging. However, Ly$\alpha$ emission originating from the diffuse gas in the filaments is theoretically predicted \citep{Hogan1987,Gould1996,Cantalupo2005,Laursen2009,Faucher2010,Goerdt2012,Rosdahl2012}.  If the only source of ionizing radiation is the UV background, which \citet{Bolton2005} and  \citet{Faucher2008} find to have a hydrogen photoionization rate $\Gamma = 0.9 (0.5) \times 10^{-12}$ at z=3 ($2\leq z \leq 4$), respectively, the Ly$\alpha$ emission from filaments is expected to be faint, on the order of SB$\simeq 10^{-20} \rm{erg  s^{-1}cm^{-2}arcsec^{-2}}$ for optically thick clouds. This is well below most modern detection limits and accounts for the large number of non-detections reported \citep{Lowenthal1990,Martinez1995,Cantalupo2005,Rauch2008}. The difficulty of detection towards lower redshifts is exacerbated by the expansion of the universe, which stretches the filaments, decreasing their density, and towards higher redshifts by the higher degree of homogeneity and less pronounced features of the cosmic web.

Even in a new era of deeper observations, a statistically robust sample of detections remains elusive. \citet{Gallego2018} stack 390 subcubes of the deepest MUSE/VLT data ($\sim$30 hrs of exposure, reaching a 2$\sigma$ surface brightness levels of $0.44\cdot10^{-20} \rm{erg s^{-1} cm^{-2}}$ in an aperture of $1''^2 \times$6.25 $\angstrom$) oriented according to the positions of Ly$\alpha$ emitters and their neighbors. They find no detectable Ly$\alpha$ emission in the intergalactic medium. 

However, when gaseous filaments are illuminated by more energetic sources such as quasars, the Ly$\alpha$ signal is boosted and detection is  made easier. Recently, Ly$\alpha$ blobs have been detected around several high-redshift quasars \citep{Martin2014,Cantalupo2014,Swinbank2015,Borisova2016,Kikuta2019}. These blobs may extend out to several hundred physical kpc and well beyond the virial radius of the host halo, suggesting that they are instead tracing the filaments of the cosmic web. When the quasars illuminating the gas are clustered, even larger extended Ly$\alpha$ nebulae (ELANe) have been detected \citep{Hennawi2015,Battaia2018}. A recent observation of filamentary Ly$\alpha$ structures between two quasars has signaled that this technique is effective for detecting more extended gaseous filaments \citep{Battaia2019}. Feedback from other energy sources, such as star formation and supermassive black hole activity may also illuminate the gas surrounding these sources above the detection threshold \citep{Umehata2019}.

Simulations have also indicated that the Ly$\alpha$ emission induced by the ultraviolet background is too faint to be currently detectable. For example, \citet{Kollmeier2010} predict that only Ly$\alpha$ emission from gas illuminated from energetic sources such as quasars should be detectable like the Ly$\alpha$ blobs mentioned above (see also \citet{Cantalupo2005}). \citet{Rosdahl2012} investigate the detectability of extended Ly$\alpha$ emission using cosmological zoom simulations and find that the Ly$\alpha$ luminosity is concentrated in the central 20$\%$ of the halo radius, making detection even more difficult. However, others are more optimistic. \citet{Bertone2012,Corlies2018,Smith2019} model Ly$\alpha$ radiative transfer in large cosmological hydrodynamical simulations to predict features of the Ly$\alpha$ emission line arising from gas in the IGM and find that MUSE and future projects such as the Square Kilometer Array (SKA) \citep{Kooistra2019},the James Webb Space Telescope and CETUS Probe Mission \citep{Hull2018} should be able to detect the brightest filaments. 
%\sgr{more background from simulation/theory work will be added. this includes relevance of future missions, an example: http://adsabs.harvard.edu/abs/2018AAS...23114013H.
%recent simulation papers:
%http://adsabs.harvard.edu/abs/2019MNRAS.484...39S ; http://adsabs.harvard.edu/abs/2018arXiv181105060C ; Bertone and Schaye 2012}.
%AMIEL Yes, in first paragraph...

In preparation for this upcoming observational revolution, our paper aims to investigate detectability of Ly$\alpha$ emission from cold gas filaments using large volume cosmological hydrodynamical simulations. In Section~\ref{sec:sims} we detail the simulations used, in Section~\ref{sec:lyaem} we describe the Ly$\alpha$ emission processes, in Section \ref{sec:maps} we focus on sources of Ly$\alpha$ emission, in Section \ref{sec:detecto} we present our main results by investigating the effects of resolution, feedback, and pixel size on the detectability of gaseous filaments, and in Section~\ref{sec:conclusions} we state the conclusions of the project.

\section{Methods}

\subsection{Simulations and Halo Selection}
\label{sec:sims} % used for referring to this section from elsewhere

\begin{table*}
\begin{tabular}{ |p{2.5cm}||p{1.8cm}|p{1.5cm}|p{1.5cm}|p{2cm}|p{1.5cm}|p{2.5cm}| }
 \hline
 \multicolumn{6}{|c|}{Simulations} \\
 \hline
 Name& Volume  &Particle &Feedback&m$_{\rm DM} $&m$_{\rm gas}$&Softening length \\
 & [(cMpc/h)$^3$] & Number & &[$\Msun$] &[$\Msun$] & [ckpc/h] \\
 \hline
 L17n1024NF &17$^3$ & $1024^3$ & No& $4.1\cdot 10^5 $   &$8.3\cdot 10^4 $& 0.39\\
 L17n512NF &17$^3$ & $512^3$ & No& $3.3\cdot 10^6 $   &$6.6\cdot 10^5 $& 0.78\\
 L25n1024NF&  25$^3$ & $1024^3$ & No& $1.3\cdot 10^6 $   &$2.6\cdot 10^5 $& 0.5\\
 L25n512NF &25$^3$ & $512^3$ & No& $1.0\cdot 10^7 $   &$2.1\cdot 10^6 $& 1\\
 L25n1024TNG  & 25$^3$& $1024^3$& Yes & $1.3\cdot 10^6 $   &$2.6\cdot 10^5 $& 0.5 \\
 L25n512TNG &25$^3$ & $512^3$ & Yes& $1.0\cdot 10^7 $   &$2.1\cdot 10^6 $& 1\\
 \hline
\end{tabular}
\caption{Description of the six simulations used. There are two volumes simulated without feedback, each at two resolution levels. For the two large volume simulations, additional simulations have been run with the IllustrisTNG feedback prescriptions.}
\label{tab:tab1}
\end{table*}

Six main simulations were used for the analysis and are described in Table~\ref{tab:tab1}. The simulations vary in resolution, volume, and feedback implementation. They are run with the moving mesh code {\sc arepo} \citep{Springel2010}, and are consistent with WMAP-9 standard cosmology, with $\Omega_m$=0.2726, $\Omega_b$=0.0456, $\Omega_\Lambda$=0.7274, and $H_0$=70.4km/s/Mpc \citep{Hinshaw2013}. Gas cooling and heating as well as star-formation follow the prescriptions given in \citet{Vogelsberger2012} and \citet{Pillepich2018}, for simulations without and with feedback, respectively. In all simulations, gas above a density threshold of $n_{\rm H}=0.13~\rm{cm}^{-3}$ follows an equation of state that is used to implicitly treat the multiphase structure of the ISM \citep{Springel2003}. Cold and dense gas above this threshold becomes eligible for star formation with a timescale that depends on the local density. In the simulations used in this analysis that include stellar and supermassive black hole feedback, denoted by 'TNG' in the simulation name, it is implemented as in IllustrisTNG \citep{Weinberger2017,Pillepich2018}. Groups are  identified based only on particle positions using the Friends of Friends algorithm. %The object sitting at the center of the gravitational potential of each halo is called the central or host galaxy and all other substructures associated with the group will be referred to as satellites or subhalos. \sgr{(this is somewhat irrelevant since later on we make no distinction between subhalos and halos and simply remove anything that is part of a halo (necessarily removing also all the subhalos). please fix accordingly the references to subhalos throughout the paper.)}

For our analysis, we choose to focus on the 12 most massive halos in the large-volume, high-resolution simulation (L25n1024NF) and the 4 most massive halos in the small-volume, high-resolution simulation (L17n1024NF) for a total of 16 halos. All halos are analyzed at redshift $z=3$. We focus on these massive halos, which span a virial mass range of $M_{\rm vir}=7.8\cdot10^{11} - 2.9\cdot10^{12}\Msun$, reasoning that larger halos may be more likely to reside in larger filaments. %\sgr{(is this really a fair statement, given that we found no dependence on halo mass within this range that we picked?)}.

\subsection{Modeling Ly$\alpha$ Emission}
\label{sec:lyaem}

%AMIEL added some words and a sentence.
Ly$\alpha$ emission can originate from two main processes: electron impact collisional excitation and radiative recombination. We assume the optically-thin limit as we are interested in detections from diffuse filaments outside of halos. In particular, we do not consider resonant scattering or dust absorption of the Ly$\alpha$ photons. One might expect this scattering to smear out bright spots in our simulated Lya maps, reducing their intensity. We have explored this possibility by performing radiative transfer on a test box of side length 1.25Mpc centered on the most massive $z=3$ halo in the simulations. The results of this validation test are summarized in Appendix A. In the case of 5.3" pixels, we find the impact of scattering to be relatively modest. The brightest pixels are reduced in intensity by a factor of ~2. 
\\
\\
$\it{Collisional}$ $\it{Excitation}$
\\
\\
%AMIEL Rephrased here
Ly$\alpha$ photons are emitted following $1s-2p$ electron impact excitation of neutral hydrogen atoms at a rate per unit volume ($\rm{cm^{-3}~s^{-1}}$) given by
%When a free electron collides with a H atom, the atom is excited to a higher energy %level before cascading back to the ground state. In the process a Ly$\alpha$ photon %is emitted. We model the Ly$\alpha$ emission rate from collisional excitation,
%$R_{\rm col}$, in units of $\rm{cm^{-3}s^{-1}}$, as:
\begin{equation}
    R_{\rm col}=q_{\rm col}(T)\cdot n_{\rm e}\cdot n_{\rm H} 
	\label{eq:col}
\end{equation}
where $q_{\rm col}$ is the collisional excitation rate coefficient, and $n_{\rm e}$ and $n_{\rm H}$ are the electron and neutral hydrogen volume number densities respectively. We adopt $q_{\rm col}(T)=(8.63\cdot10^{-6}{\rm cm^3 s^{-1}})\cdot e^{-E_{\rm Ly\alpha}/{\rm k}T}/(2\sqrt{T})$ \citep{Dijkstra2017} where $T$ is the gas temperature in Kelvin, $E_{\rm Ly\alpha}$=10.19 eV is the energy of the Ly$\alpha$ transition, and $k$ is the Boltzmann constant.
\\
\\
$\it{Recombination}$
\\
\\
%AMIEL are you assumin case A or case B.  If case A you need to use an effective recombination coefficient, not the total. I don't understand what you mean by "normalization".
For radiative recombination we assume the optically thin case A, for which the Ly$\alpha$ emission rate per unit volume is 
%every electron prototn recombination leads to the 
%When a free electron combines with a free proton, it cascades from a high energy %level to a lower, more stable energy level, possibly emitting a Ly$\alpha$ photon. %The Ly$\alpha$ emission rate from recombination, $R_{\rm rec}$, in units of %$\rm{cm^{-3}s^{-1}}$, is modeled here as:
\begin{equation}
    R_{\rm rec}=\alpha(T)\cdot n_{\rm e}\cdot n_{\rm p} 
	\label{eq:rec}
\end{equation}
where $\alpha(T)$ is the effective Ly$\alpha$ case A recombination coefficient, $n_{\rm p}$ is the proton volume number density and $n_{\rm e}$ is the electron volume number density. For Ly$\alpha$ emission, $\alpha(T)=\alpha_{\rm tot}(T)\cdot f_{\rm rec}(T)$ where $\alpha_{\rm tot}$ is the total case A recombination coefficient for HII and $f_{\rm rec}$ is the fraction of recombination that results in emission of a Ly$\alpha$ photon. We adopt from \citet{Hui1997}, 
\begin{equation}
    \alpha_{\rm tot}(\rm T)=1.269\times10^{-13}\cdot\frac{\lambda^{1.503}}{(1+(\lambda/0.522)^{0.7})^{1.923}} \rm{cm^3 s^{-1}},
    \label{eq:atot}
\end{equation} 
where $\lambda=2\cdot157807/T$, and from \citet{Dijkstra2017},
\begin{equation}
   f_{\rm rec}(T)=0.41-0.165\cdot  {\rm log}_{10}(T/10^{4})-0.015\cdot(T/10^{4})^{-0.44}.
   \label{eq:frec}
\end{equation}
 Eq.~\ref{eq:frec} is a fit to a non-analytic expression of f$_{rec}$ for Case A that is valid out to T=4$\times 10^5$ K (see Eq.15 in \citet{Dijkstra2017}). Therefore, beyond this temperature we adopt a constant $f_{\rm rec}(T > 4\times10^5{\rm K}) = f_{\rm rec}(4\times10^5{\rm K})=0.14$. 
 %\sgr{have we not decided to use the fit all the way up to $4\times10^5$, which is the actual range for which the fit was made? also, is eq. 4 not a fit to the Dijkstra calculations, shown in their solid line, which also indeed fit to the more limited range of data from osterbrock? if i am not mistaken, saying that eq. 4 is a fit to osterbrock is somewhat misleading.}

\begin{figure}
	\includegraphics[width=\columnwidth]{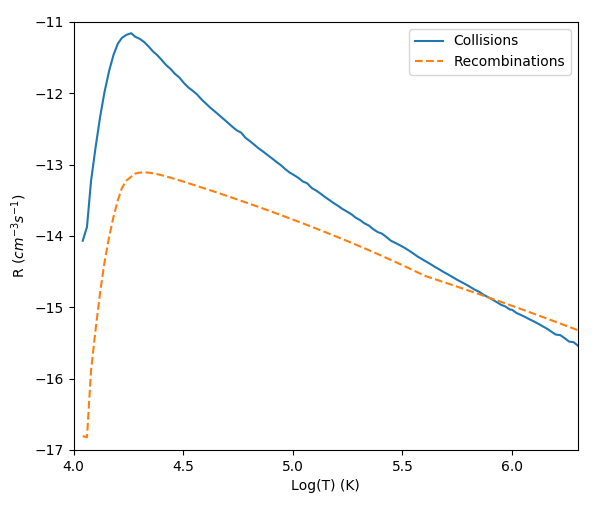}
    \caption{Temperature dependence of collisional excitation and recombination rates for hydrogen gas Ly$\alpha$ emission in collisional ionization equilibrium. Rate is per unit density squared.}
    \label{fig:js}
\end{figure}

%AMIEL Please remind me, when you made Fig 1 did you compute the ionization state of the gas? If so, this has to be explained!
The temperature dependence of collisional and recombination emissivity is shown in Fig.~\ref{fig:js} assuming the hydrogen is in collisional ionization equilibrium and ignoring photoionization (e.g \citet{Gnat2007}). At higher temperatures, the gas becomes completely ionized and recombinations dominate, while at low temperatures the neutral fraction increases and collisional excitation dominates. The total Ly$\alpha$ luminosity from every simulation gas cell (without photoionizations), in units of $\rm{erg/s}$, equals $L_{\rm tot} \equiv (R_{\rm col}+R_{\rm rec})\cdot E_{\rm Ly\alpha}\cdot V$, where $V$ is the volume of the cell. 

%\setlength{\belowcaptionskip}{-25pt}
%\begin{center} 
\begin{figure*}
    \centering
	\includegraphics[width=2\columnwidth]{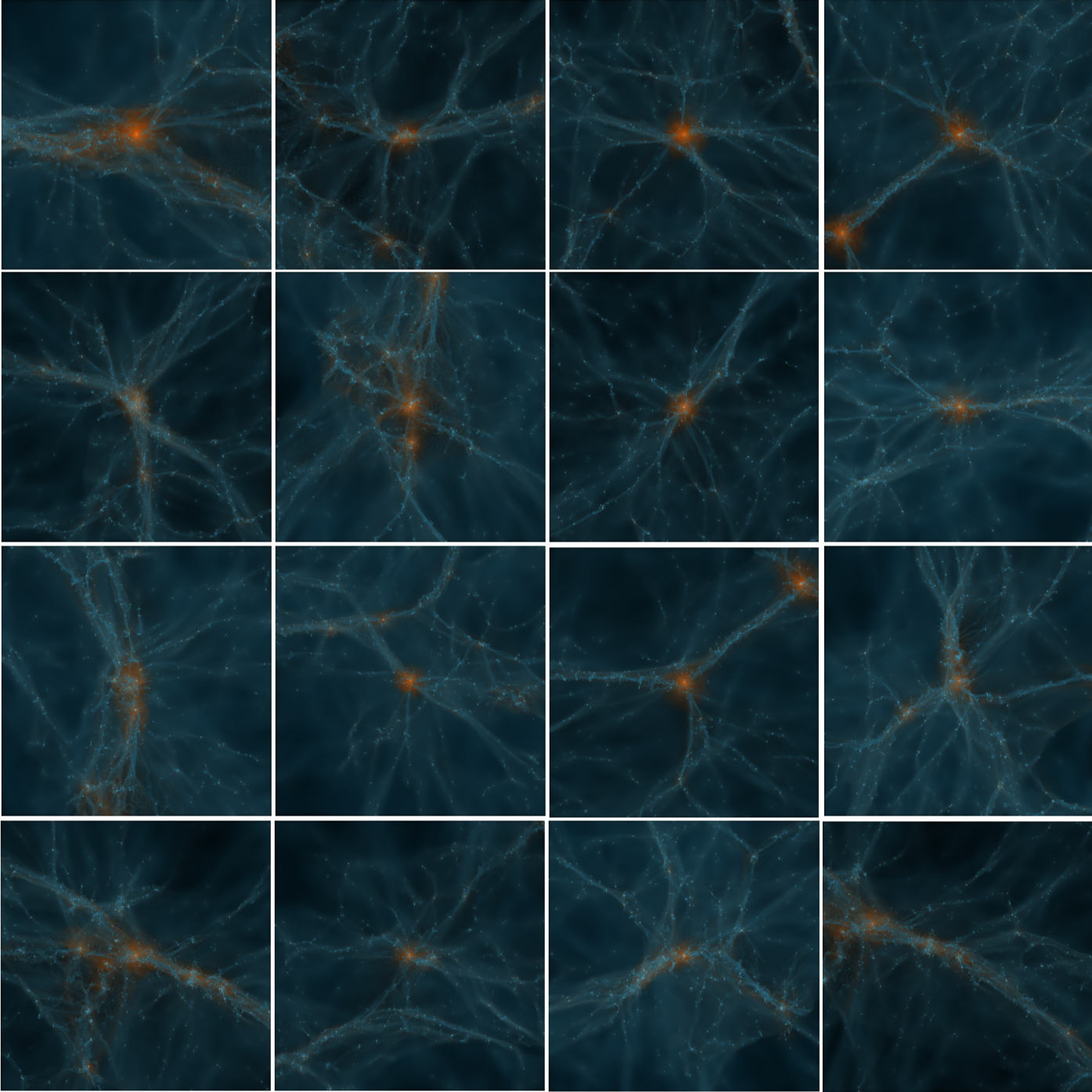}
    \caption{Temperature and density maps of $(5 \rm{cMpc})^2$ regions centered on each halo from simulations L25n1024NF and L17n1024NF. Density is shown in blue and temperature in orange. First three rows show the four most massive halos (ordered in mass from most massive at top left to least massive on the bottom right) in the high resolution, $(25 {\rm cMpc/h})^3$ box. The bottom row shows most massive halos in the high resolution, $(17 {\rm cMpc/h})^3$ box. Filaments are populated by high temperature and density subhalos, which we remove in the remaining analysis to focus on the cold, gaseous filaments.}
    \label{fig:15ims}
\end{figure*}
%\end{center}
\subsection{Image Generation}
\label{subsec:image}
 To create our final surface brightness images of filaments we first remove any particles identified by FoF to be within a halo, both to avoid regions that are expected to be optically-thick and because we do not resolve the multi-phase structure of the interstellar medium inside galaxies. Since we are focused on exploring the detection of the cosmic web outside of dark matter halos, this excision is not expected to affect our results.

We then cut out a $(5~\rm{cMpc})^2$ comoving region ($(1250 {\rm kpc})^2$ in physical units) centered on each halo. In the `line of sight' direction (which we arbitrarily choose to be the complete z axis of the simulation box) we make a velocity-based cut such that only gas cells with galacto-centric velocities in a range of $\pm (\Delta v/2=193$ km$\rm s^{-1})$ are included. This velocity range is chosen to be consistent with the narrow band wavelength width of the images in \citet{Gallego2018}, which are composed of 5 stacked voxels, each of width 1.25 $\angstrom$ to reach a total wavelength width of $\Delta \lambda=6.25$ $\angstrom$. We convert the wavelength width into a velocity range via Eq.~\ref{eq:doppler}: 
\begin{equation}
    \Delta v = {\rm c} \cdot \frac{\Delta \lambda}{\lambda(1+z)}
	\label{eq:doppler}
\end{equation}
where c is the speed of light, $\Delta \lambda$ is the wavelength width of the stacked image, and $\lambda$ is the Ly$\alpha$ rest wavelength. This volume shows the extent of the filaments while not being so large so as to increase the likelihood of contamination from other massive centrals. 
  %\begin{center}
\begin{figure*}
    \centering
    \vspace{-0.3cm}
	\includegraphics[width=2\columnwidth]{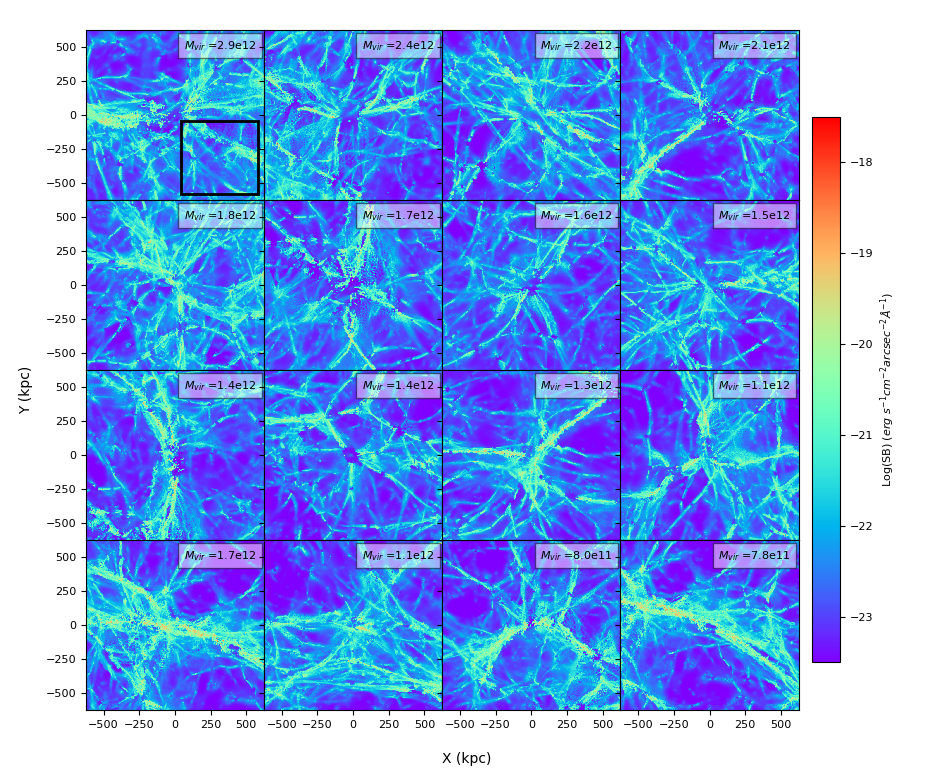}
    \caption{Pseudo narrow-band images of Ly$\alpha$ surface brightness of the regions shown in Fig.~\ref{fig:15ims}, which at $z=3$ are 2.6 arcmin on a side, created using L25n1024NF and L17n1024NF. All subhalos are excised and the pixel size matches that of MUSE. Box in top left panel indicates MUSE Hubble Ultra Deep Field exposure area. In comparison with Fig.~\ref{fig:15ims}, the locations of brightest Ly$\alpha$ emission correspond to the locations of high temperature and density.}
    \label{fig:15jims}
\end{figure*}
%\end{center}
Since we have removed subhalos from our analysis, all surface brightnesses reported here are lower limits for detection. We create maps of Ly$\alpha$ surface brightness (Fig.~\ref{fig:15jims}) by summing the Ly$\alpha$ signal along the 6.25$\angstrom$ width in the spectral or z direction and dividing by 6.25$\angstrom$  so as to obtain an averaged flux density in units of $\rm{erg\,s^{-1}cm^{-2}arcsec^{-2}\angstrom^{-1}}$.
\\
\\
\\
\\
Finally, in order to account for the numerical effects of particle resolution, we use the smoothing code {\tt swiftsimio}\footnote{\url{https://pypi.org/project/swiftsimio/}} to smooth the gas particles with a 2D Wendland C2 kernel whose FWHM is equal to the distance to the 15th nearest neighboring gas particle. We choose 15 for the number of neighbors as it is the typical number of faces of a Voronoi cell given a pseudo-random distribution of mesh-generating points in 3D. To test the validity of this choice, we also try smoothing with kernel FWHM values equal to the distance to the 30th and 60th neighbor. We find that the results are converged, and therefore adopt the distance to 15th nearest neighbor as the FWHM of the smoothing kernel for the remainder of the analysis.

\subsection{Detection Limit}
\label{subsec:detlim}
  MUSE is a panoramic, integral-field spectograph currently operating on the Very Large Telescope (VLT) of the European Southern Observatory (ESO). The MUSE Hubble Ultra Deep Field survey includes an ultra deep, 30.8 hour exposure of a 1.15 arcmin$^2$ field. The field has a propagated noise standard deviation of $0.33\cdot10^{-20} \rm{erg s^{-1} cm^{-2}\angstrom^{-1}}$ for a sampling voxel. In Wide Field Mode, MUSE has a spatial sampling of 0.2"$\times$0.2" and a spectral line resolution of 1.25$\angstrom$. %\sgr{(i'm not sure... isn't figure 18 the histogram of values of native pixels in empty regions, i.e. pixels of 1.25 in the spectral direction, not 2.5, and 0.2'' on a side, not 1''?)}
  \citep{Bacon2017}\footnote{Note the inconsistency between the values shown in Figures 18 and 19 of \citet{Bacon2017}, which is a result of an incorrect conversion of the raw measurements shown in Figure 18 into the quantity shown in Figure 19 (Bacon R., private communication). In this work, we use only the width of the histogram shown in their Figure 18 to derive the appropriate surface brightness limit directly from the pixel noise.}. Voxels in our analysis have the same area as the sampling voxels, but a spectral line width of 6.25$\angstrom$ (consistent with \citet{Gallego2018}). To obtain the surface brightness limit for our voxels, we first convert the noise to units of $\rm{erg s^{-1} cm^{-2}}$:
  $0.33\cdot10^{-20}\rm{erg s^{-1} cm^{-2}\angstrom^{-1}}\times1.25\angstrom=0.4125\times10^{-20}\rm{erg s^{-1} cm^{-2}}.$ %Therefore, the surface brightness emission line sensitivity for this sampling-size voxel is $0.33\cdot10^{-20}/0.2^2 8.25\cdot10^{-20} \rm{erg s^{-1} cm^{-2}arcsec^{-2}\angstrom^{-1}}. 

  Our voxels have the equivalent depth of 5 Wide Field Mode sampling voxels. Since noise scales as the square root of the number of voxels, the noise for a single voxel of area 0.2"$\times$0.2" and depth 6.25$\angstrom$ level is:
  %\begin{equation*}
    \begin{flalign}
\sigma &= 0.4125\cdot10^{-20}\times \sqrt{5}\\\nonumber
&=9.2\cdot10^{-21}\;\rm{erg\,s^{-1}cm^{-2}}
    \end{flalign}

    Finally we convert this value to units of surface brightness. Our surface brightness detection limit for a Wide Field Mode sampling voxel at the 5$\sigma$ level is:
     \begin{flalign}
{\rm SB}_{\rm lim} &= 5\times9.2\cdot10^{-21}\rm{erg\,s^{-1}cm^{-2}}/6.25\angstrom/0.2^{2}arcsec^2\\\nonumber
&=1.84\cdot10^{-19}\;\rm{erg\,s^{-1}cm^{-2}arcsec^{-2}\angstrom^{-1}}
    \end{flalign}
  %\end{equation*}

\section{Results: Sources of L\lowercase{y}$\alpha$}
\label{sec:maps}
Images of the halos in temperature and density are shown in Fig.~\ref{fig:15ims}, and in total Ly$\alpha$ surface brightness in Fig.~\ref{fig:15jims}. The areas of high density (shown in light blue) in Fig.~\ref{fig:15ims} correspond to the highest Ly$\alpha$ emission in Fig.~\ref{fig:15jims}. Areas of high temperature (shown in orange) in Fig.~\ref{fig:15ims} are mostly confined to halos which are excised in Fig.~\ref{fig:15jims} and thereafter, as described in Section \ref{sec:lyaem}.

%\begin{center}
\begin{figure}
	\includegraphics[width=0.48\textwidth]{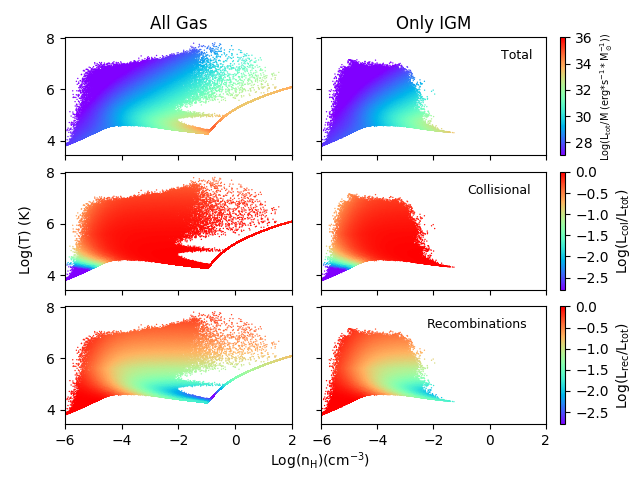}
    \caption{Temperature vs. density phase space diagrams for gas cells within a region of (1250 kpc)$^2$ and velocity depth of v= $\pm$ 193 km/s for the most massive halo in the L25n1024NF simulation (top left panel of Fig.~\ref{fig:15jims} and \ref{fig:15ims}). The left column includes particles bound to subhalos, the right column does not. From top to bottom: phase space colored by total Ly$\alpha$ emission, fraction of emission due to collisions, and fraction of emission due to recombinations. In regions of high density and low temperature, collisional excitation dominates as expected from Fig.~\ref{fig:js}.}
    \label{fig:phase}
\end{figure}
%\end{center}

To explore the dependence of the total Ly$\alpha$ luminosity on the physical conditions of the gas and discern the contributions of collisional excitation vs. recombination, we investigate the phase diagram of the gas within the spatial and velocity bounds mentioned above. Since in this context the results are essentially invariant between different halos, we focus here on the most massive halo, the one that is shown in the top left panel of Fig.~\ref{fig:15ims} and Fig.~\ref{fig:15jims}. We begin with Fig.~\ref{fig:phase}, which presents the intrinsic dependence of the luminosity on density and temperature irrespective of the actual mass distribution of gas in the phase diagram. The left column was created including subhalos while for the right column the subhalos have been excised. The color in the first row indicates the total Ly$\alpha$ emission (from both processes) per unit mass as a function of gas density and temperature. The second row is colored by the fraction of total emission due to collisional excitation, and the third by the fraction of total emission due to recombinations. Until this point we have assumed collisional ionization equilibrium, which only considers collision and recombination processes. The luminosity from both is only dependent on temperature. However, in the simulation, photoionizations are also possible, which creates an implicit dependence of ionization fraction on density. The ionization fraction of the gas decreases with density and increases with temperature, reducing $n_{\rm H}$ and increasing $n_{\rm e}$ and $n_{\rm p}$. This relation is reflected in Fig.~\ref{fig:phase} which shows recombination dominating the total emission at low densities and high temperatures, while the contribution of collisional excitation is largest at low temperatures. Above $n_{\rm p}+n_{\rm H} \simeq 10^{-2.5} \rm{cm^{-3}}$ almost all emission is due to collisional excitation .
%AMIEL rephrased here, maybe should move this to a footnote.
\footnote{The `tail' feature in the plots in the left column reflects the polytropic equation of state assumed in the simulations to prevent artificial fragmentation at high densities 
%where we have insufficient resolution to follow the relevant physics;
\citep{Springel2003}.} 

%\begin{center}
\begin{figure}
	\includegraphics[width=\columnwidth,height=6.5cm]{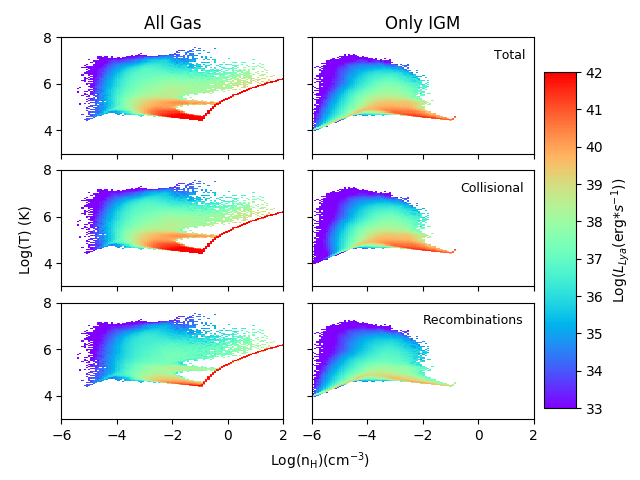}
    \caption{Density of phase space for gas cells within a region of (1250 kpc)$^2$ and velocity depth of v= $\pm$ 193 km/s. Median of all 16 halos. Left column includes particles bound to halos, right column does not. From top to bottom: 2d histograms weighted by total Ly$\alpha$ emission, emission due to collisions, and emission due to recombinations. 100x100 pixels in each panel. Emission from collisions dominates total emission.}%\sgr{see comments about previous figure. in addition: for the units on the colorbar to make sense, the pixel size (on this T-density plane) has to be defined, isn't that the case? if so, please quote what it is.}}
    \label{fig:densphase}
\end{figure}
%\end{center}

From Fig.~\ref{fig:phase} alone, it is impossible to tell which process is responsible for the majority of $\textit{total}$ emission that is potentially detectable, since it presents the luminosity per unit mass. 
\begin{center}
\begin{figure*}
	\includegraphics[width=2\columnwidth]{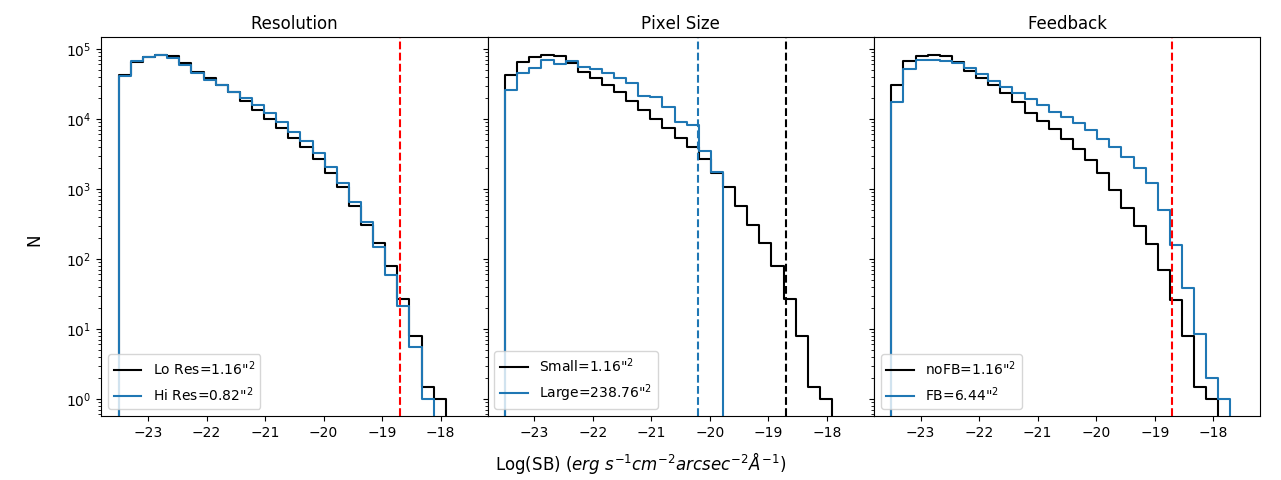}
   \caption{Effects of changing simulation resolution, pixel size, and feedback on surface brightness. Each histogram represents the median distribution of the total Ly$\alpha$ surface brightness of the 16 galaxies in Fig.~\ref{fig:15ims}. Black (blue) histogram corresponds to low (high) resolution in the left panel, small (large) pixel size in the middle panel and no feedback (feedback) in the right panel, respectively. The dashed red line shows the MUSE detection limit. In the middle panel, there are two detection limits shown since the detection limit varies with pixel size and the blue histogram is renormalized so that the area below the histogram matches the others. Numbers in bottom left corner denote detectable area on the various types of average images, each of which has a total area of $(2.64')^2=25000''^2$. Simulation resolution does not significantly affect the detectable area. However, larger pixel size and implementing feedback both significantly increase detectability.}
    \label{fig:threemeds}
\end{figure*}
\end{center} 
Fig.~\ref{fig:densphase} addresses this %\sgr{(i don't think it really addresses this because it knows nothing about the detection limit of MUSE or any other instrument, and it says nothing about surface densities...)}
by showing the distributions in phase space of total, collisional excitation, and recombination emission (from top to bottom), averaged over all 16 halos in Fig.~\ref{fig:15ims} %\sgr{(i think better to use all halos combined for this, because this can vary from one system to the other and we don't a-priori know by how much or whether this one particular halo in biased or not)} 
and Fig.~\ref{fig:15jims}.  The resulting areas of high and low emission are a reflection of both the luminosity per unit mass and the mass distribution on the phase diagram.Fig.~\ref{fig:densphase} clearly shows the largest total amount of Ly$\alpha$ emission %\sgr{(this is not surface density, this is emission per pixel on the T-density plane, so we need to use more precise language than "highest")}
is due to gas with temperature below $T\simeq 10^5$K and density above $n\simeq 10^{-3} \rm cm^{-3}$. At these temperatures and densities, collisional excitation dominates over recombination (see Fig.~\ref{fig:phase}, second row). Although detection of Ly$\alpha$ emission is dependent on the spatial distribution of the gas, our results suggest that collisional excitation is a more likely source of detectable emission than recombination. In fact, it is responsible for 95$\%$ of the total emission, on average, assuming no additional sources of photoionization. \footnote{We note that this result is unchanged whether the analysis is performed with data from the low- or high-resolution simulation.}

\section{Results: Improving Detectability}
\label{sec:detecto}
We now focus on the detectability of the gaseous filaments surrounding massive halos with the MUSE instrument. We describe the effect of varying simulation resolution (Section \ref{subsec:simres}), pixel resolution (Section \ref{subsec:pixres}), and feedback models (Section \ref{subsec:fdback}) in order to optimize detection.

\subsection{Simulation Resolution}
\label{subsec:simres}
A natural step towards boosting the signal is to increase the resolution of the simulation to reveal substructure within the filaments. Since the emissivity scales as density squared, small variations of the gas in phase space could be amplified in Ly$\alpha$ emission. Gas at higher densities will increase the total Ly$\alpha$ emission, making the filament more detectable.

To quantify the expected signal boost, the left panel of Fig.~\ref{fig:threemeds} compares the median distribution of total Ly$\alpha$ surface brightness per pixel for all 16 halos from the low (black) and high (blue) resolution simulations with no feedback (L25n512NF and L25n1024NF, respectively). The red vertical line represents the MUSE detection limit, while the number in the bottom left corners represent the median combined angular area of pixels above the MUSE detection limit. 

The size of the MUSE pixel is $\sim$1.5 kpc on a side - small enough to probe fine substructure. It is therefore surprising that at most surface brightness levels, the images created with higher resolution simulations are consistent with their low resolution counterparts, suggesting that any overdensities that are resolved by increasing resolution are not significant enough to boost the Ly$\alpha$ surface brightness. Small differences occur between the two simulation resolutions at mid-range surface brightnesses (Log(SB)$\sim$-21) and at the high surface brightness tail.In this regime, which represents the detectable pixels, the images created with a low resolution simulation have a factor of $\simeq$1.5 more detectable area than images from the high resolution simulation, on average. The differences between the two resolutions are a numerical artefact of the low resolution creating dense `clumps' of particles that are spread more continuously over neighboring pixels at higher resolution. The numerical effect is dependent on both pixel resolution and feedback, which will be discussed in Sec.~\ref{subsec:pixres} and  Sec.~\ref{subsec:fdback}, respectively. Observations, of course, are free from numerical resolution effects. Overall our findings suggest, somewhat counterintuitively, that the degree of substructure within the filaments is not a strong indicator of their detectability.

\subsection{Pixel Resolution}
\label{subsec:pixres}
 Observationally, detecting filaments has proven difficult and has often relied on stacking techniques \citep{Steidel2000,Matsuda2004,Matsuda2011}. Stacking images containing filaments is challenging to do observationally because it requires the images to be stacked in such a way that the filaments align, but their geometry is not known a priori (e.g.~\citealp{Steidel2010}). \citet{Gallego2018} attempt to overcome this limitation by using positions of nearby Ly$\alpha$ galaxies and assuming they are connected by a filament to orient subcubes before stacking, but are not able to make a detection even with 390 `oriented' stacks of MUSE data. They thus suggest that 2/3 of their cubes may not contain filaments. 
 
We hence turn to a different image processing technique: pixel degradation. The effects of artificially resizing pixels on detectability have been addressed in similar contexts \citep[see, e.g][]{Bertone2012, Hani2020}. We note that \citet{Gallego2018} effectively use pixel degradation in their analysis by 'resampling' so that the area of their pixels increases by a factor of two in order to avoid empty ones. Here we suggest a more extreme resampling. By decreasing the number of pixels in an image (effectively combining them), the number of Ly$\alpha$ photons per pixel is increased, thus increasing the signal by a factor of $n$, where $n$ is the factor by which the angular area of a pixel is increased, and the noise by a factor of $\sqrt{n}$, thereby increasing the signal-to-noise ratio by a factor of $\sqrt{n}$. The drawback to this technique is that the continuum of the sources needs to be removed to high accuracy, and of course the lost information on smaller substructures within the filaments. However, if our main aim is simply to detect a filament, degrading is sufficient for this purpose. 
 
 \begin{figure}
	\includegraphics[width=0.5\textwidth]{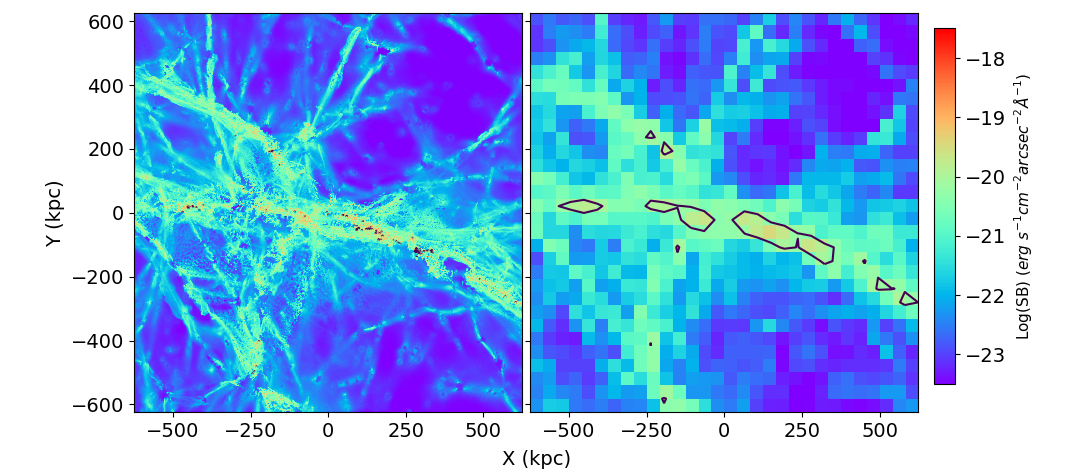}
    \caption{MUSE pixel size vs. degraded pixel size images of the galaxy in the bottom left of Fig.~\ref{fig:15ims} and Fig.~\ref{fig:15jims} from the L17n1024NF simulation. Black contours show MUSE detection limit. In the left (right) panel there are 791 (30) pixels on a side, resulting in a pixel area of (0.2")$^2$ ((5.3")$^2$). All filaments are more easily detected at the larger pixel size.}
    \label{fig:degmock}
\end{figure}

Fig.~\ref{fig:degmock} shows an example for the results of the degrading technique. The area of each pixel in the right panel is $\simeq 700$ larger than the area of each MUSE-like pixel in the left panel. Changing the pixel size has also changed the detection limit, which scales as $1/\sqrt{n}$ where n is the ratio of the area of the pixel to the area of a pixel of MUSE sampling size. Several pixels are now above the threshold for detection by MUSE, and begin to delineate the filaments. For example, the filament that extends below the main horizontal one is now detected, whereas with the MUSE pixel size an observer would never have known of its presence. In order to maximize the effectiveness of this technique, the area of the pixel should be chosen to minimize contamination from nearby pixels with low S/N values. We have therefore chosen the size of pixel to roughly match the width of a typical filament (not the largest filament). 

The middle panel of Fig.~\ref{fig:threemeds} better quantifies the effect of pixel degradation by binning the Ly$\alpha$ luminosities in images of both pixel sizes. Blue and black histograms are derived from images of pixel area $(5.3)"^2$ and $(0.2)"^2$, respectively, and dashed lines illustrate corresponding detection limits. The total detected image area is clearly greater when using large pixels. On average, the detectable area is over two orders of magnitude larger for the degraded images. However, in most cases, $\textit{more}$ pixels are detectable with a small pixel size. For example, for the second-most massive galaxy (top row, second from left galaxy in Fig.~\ref{fig:15jims}), only one large pixel is detected, compared to 4 small pixels. This suggests that either all of the small detected pixels are located in an area encompassed by the large detected pixel, or, far more likely, they are more spread out but the degree of degradation is too large, such that when they are combined with nearby pixels of lower luminosities, potentially located in a void, the signal is diluted. However, in most cases, such as for the halo in the bottom left of Fig.~\ref{fig:15jims}, combining pixels allows for a clear detection. In this case, instead of detecting many smaller pixels that may lie in several different filaments, pixel degradation facilitates detection of fewer, larger pixels that trace filaments more reliably. Our technique is not new, however, to our knowledge, this is the first time degrading to such an extent has been suggested for use on deep MUSE data.

\subsection{Feedback}
\label{subsec:fdback}

Until this point all results have been derived using simulations without feedback. We now investigate the effects of feedback on the detectability of filaments using the identical initial conditions but with supernovae and black hole feedback modeled in the same way as in IllustrisTNG. The right panel of Fig.~\ref{fig:threemeds} compares the average total Ly$\alpha$ surface brightness histograms for the 12 halos from the low resolution, (25cMpc/h)$^3$ box simulation with (blue, L25n512TNG) and without (black, L25n512NF) feedback. The four halos from the high resolution, (17cMpc/h)$^3$ box are not included as we do not have available an equivalent simulation with feedback. 

The right panel of Fig.~\ref{fig:threemeds} clearly demonstrates the strong effect of feedback on detectability. We find that feedback increases the number of detected pixels by a factor of $\simeq$ 5.5 on average. That such a substantial signal boost should occur is not obvious. Black hole-driven winds have been shown to destroy galactic disks and other structures \citep{Beckmann2019,Grand2017}. Gas filaments, which have low densities and are not bound, are especially susceptible to disruption through feedback processes. However, we find that the amount of gas that is blown out of the filaments due to feedback is negligible when compared with the amount of gas blown from the central halo into them.

\begin{figure}
	\includegraphics[width=0.5\textwidth]{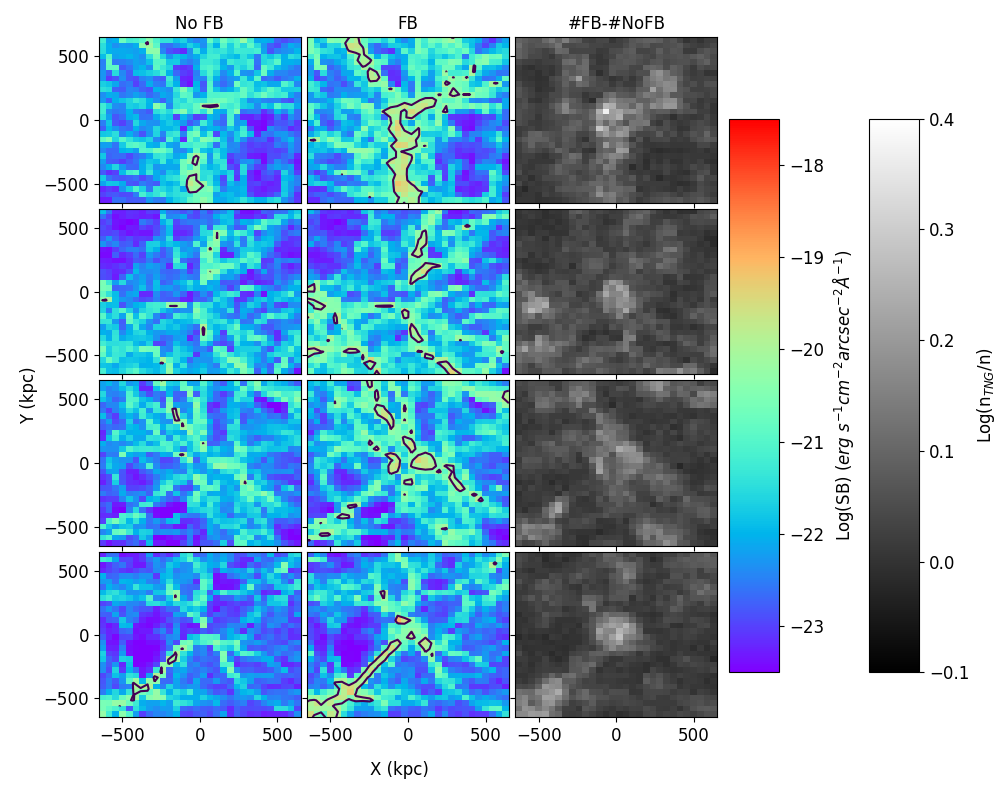}
    \caption{First column: Degraded images of the four most massive galaxies in the sample (shown in the top row of Fig.~\ref{fig:15ims} and Fig.~\ref{fig:15jims}) created using the high resolution, large volume simulation without feedback (L25n1024NF). There are 30 pixels on a side, resulting in a pixel area of (5.3")$^2$. Second column: Same as the first column but with feedback (L25n1024TNG). Black contours denote MUSE detection limit. Third column: Enhancement in number of gas cells per pixel when implementing feedback in the low resolution simulation. Enhanced pixels coincide spatially with filaments.}
    \label{fig:densenh}
\end{figure}

Fig.~\ref{fig:densenh} shows the increase in number of gas cells in each pixel between the feedback and no-feedback simulations, coupled with the effect of pixel degradation, for the four halos in the top row of Fig.~\ref{fig:15ims} and Fig,~\ref{fig:15jims}. The third column illustrates the difference between the number of gas cells occupying each pixel in the images of the first and second columns. Boosts in number of gas cells (shown in white) are preferentially located along existing filaments, while voids experience almost no change. This is consistent with our understanding of the feedback model. Filaments are populated by small halos whose galactic winds travel at velocities that scale with local dark matter velocity dispersion \citep{Pillepich2018}. The wind velocities do not allow gas to easily escape their potential wells. Thus, most feedback is localized to the halo, while any gas that escapes the halo is not likely to have a high enough velocity to escape the filament. Therefore, although we excise the halos, traces of feedback are mostly confined to the filaments.

The TNG feedback model is so effective that filaments are detectable even at the MUSE pixel size. That we have not yet observed a filament with MUSE suggests that the model is ejecting too much gas from subhalos into filaments. One possible reason for this could be the speed of the galactic winds. The TNG feedback model relies on strong ejective feedback instead of preventative feedback to regulate gas. For low mass galaxies, the TNG model predominantly consists of galactic winds that are faster, warmer, and more efficient than the previous Illustris model \citep{Pillepich2018}. The speeds of the TNG galactic winds are fast enough to escape the halo, but not the filament it resides in, resulting in an excess of gas within the filaments. While this results in a better match to observational constraints such as galaxy size and stellar content at the low mass end, it may not lead to a realistic gas distribution at larger radii.

\section{Conclusions}
\label{sec:conclusions}
Our standard cosmological model, $\Lambda$CDM, predicts that gas in the intergalactic medium is distributed in filaments linking massive halos. Furthermore, cosmological simulations suggest that these filaments trace underlying dark matter filaments that make up the cosmic web. Recombinations and collisional excitation processes result in faint Ly$\alpha$ emission from the gaseous filaments, allowing for possible detection by deep imaging instruments such as MUSE. In this study we have used six simulations of differing resolutions and physics models to investigate the detectability of the filaments feeding 16 massive ($M_{\rm vir}= 7.8\cdot10^{11} - 2.9\cdot10^{12}\Msun$) halos  at z=3 with the MUSE/VLT instrument. We summarize our findings below:

\begin{itemize}

\item Assuming an optically thin limit, we map total Ly$\alpha$ emission (collisional excitation and recombination) and find it correlates  with the temperature and density of the filaments as expected. That is, collisional excitation (recombination) channels dominate at high (low) densities and low (high) temperatures.

\item The vast majority of the Ly$\alpha$ emission from gas  (excluding that in halos) is caused by collisional excitations which dominate in low temperature, high density regions. It is not entirely surprising that most emission arises from cool clumps instead of hot, ionized streams of gas. \citet{Furlanetto2005} find a similar domination of collisonal processes in a similar redshift, temperature, and density regime. \citet{Witstok2019} analyze Ly$\alpha$ emission at z=5.76 using the Sherwood simulation suite and, assuming an optically thin limit, find that while collisional processes dominate in high density areas, recombinations dominate at lower densities common in filaments. Finally, \citet{Rosdahl2012} compute Ly$\alpha$ emission from filaments surrounding a halo of the same virial mass as the most massive halo in this study (2.9$\times10^{12} M_\odot$) and find that ~95$\%$ of the total Ly$\alpha$ emission is contributed through collisional channels. However, we note that their analysis focuses on the inner regions of the halo (r<r$_{vir}$), thus excluding the gas from the long, extended filaments. Furthermore, we have removed any particles associated to any halo via {\sc subfind}, while \citet{Rosdahl2012} employ radiative transfer and self-shielding models to these regions. However, it is worth noting that for observations, additional sources of photoionization could enhance recombination emission, especially at large distances from dense subhalos.

\item Changing the resolution of the simulation does not result in a significant signal boost. This suggests that filaments with a higher degree of substructure will not be significantly more detectable.

\item By degrading the pixel resolution of the original mock image by a factor of $\simeq 700$ in pixel area, the area of the image above the detection limit is boosted by a factor of $\simeq 200$ on average. Although we have lost information about small scale structure we are easily able to detect diffuse gaseous filaments at $z=3$ to the 5$\sigma$ level.

\item Implementing feedback increases the detectable area by a factor of $\simeq 5.5$ on average in filaments. Furthermore, the effect is focused along the filaments: gas is blown out from the halos lying along the filaments, increasing filament densities and thus boosting Ly$\alpha$ emission. When combined with the effect of pixel degradation, filaments are clearly detected. Even using the MUSE pixel size, employing the TNG feedback model boosts a considerable number of pixels beyond the detection threshold, suggesting that filaments should be observable. The fact that we have not observed them yet indicates that the TNG model is most likely  overpopulating the filaments with gas.

\end{itemize}

%Application of our degradation technique on recent and future MUSE data is an exciting avenue for the possible detection of filaments comprising the cosmic web. %Future work could also involve employing Voronoi mesh methods will as an alternative to our simpler two-dimensional binning technique in order to create mock images and potentially boost signal, given their superior sensitivity to density. 

\section*{Acknowledgements}
We would like to thank the referee for improving our work with their helpful comments and insights. This work was initiated as a project for the Kavli Summer Program in Astrophysics held at the Center for Computational Astrophysics of the Flatiron Institute in 2018. The program was co-funded by the Kavli Foundation and the Simons Foundation. We thank them for their generous support. AS thanks the German Science Foundation for support via DFG/DIP grant STE 1869/2-1 GE/625-17-1 at Tel Aviv University.

%%%%%%%%%%%%%%%%%%%%%%%%%%%%%%%%%%%%%%%%%%%%%%%%%%



%\bibliographystyle{mnras}


%%%%%%%%%%%%%%%%%%%%%%%%%%%%%%%%%%%%%%%%%%%%%%%%%%

%%%%%%%%%%%%%%%%% APPENDICES %%%%%%%%%%%%%%%%%%%%%
%\clearpage
\appendix
\section*{Appendix A}

To compute the impact of Ly$\alpha$ scattering, we have used a parallelized 3D Monte Carlo Ly$\alpha$ radiative transfer code, which was originally developed for line-intensity mapping predictions \citep{Visbal2018}. This code is similar to the one described in \cite{Faucher2010} \citep[see also][]{Zheng2002, Cantalupo2005, Dijkstra2006, Laursen2007}. For a detailed description see Appendix C in \citet{Faucher2010}.  To generate a distant-observer image, the code follows the paths of Ly$\alpha$ photons packets as they scatter throughout the simulated neutral gas field. 

For the radiative transfer calculation, the neutral hydrogen density, temperature, Ly$\alpha$ emissivity, and gas velocity in a (1250 kpc)$^3$ volume centered on our most massive galaxy (top left in Fig.~\ref{fig:15ims} and Fig.~\ref{fig:15jims}) were smoothed over a $500^3$ cubic grid using a gaussian kernel whose width depends on the local density such that it is smaller in denser regions. We then follow $\sim 10^6$ Ly$\alpha$ photons packets, distributed proportionally to the filaments' Ly$\alpha$ emission to estimate the distant-observer image cube (two spatial dimensions and one frequency dimensions). We also utilize the ``accelerated scheme'' described in Appendix C of \cite{Faucher2010} with $x_{\rm crit}=2.5$, which greatly speeds up our calculation without significantly impacting the results. 

We have checked convergence for the number of packets, $x_{\rm crit}$, the grid resolution, and the integration of packets' paths through the simulation. Our results are summarized in Fig.A1 where we plot the histogram of observed Ly$\alpha$ intensities with and without performing radiative transfer. Here we have convolved our $500^3$ resolution image cube with a top-hat window function with dimensions to 5.3" x 5.3" x 6.25 Angstroms. This corresponds to the degraded resolution used in Section \ref{subsec:pixres}. We see that the smearing of the signal due to radiative transfer reduces the intensity of the brightest pixels by a factor of $\sim 2-3$ compared to ignoring this effect. Shifting the surface brightness distributions in Fig.~\ref{fig:threemeds} lower by a factor of $\sim 2-3$ would alter the detectable area, particularly for the large pixel size, which has a steep cutoff at the high surface brightness end. We have not checked how radiative transfer affects surface brightness at different pixel sizes, simulation resolutions, or feedback implementations. Given the above calculations, we expect the impact to be relatively modest (particularly in the large pixel size case).  

\renewcommand{\thefigure}{A\arabic{figure}}
\setcounter{figure}{0}
\begin{figure}
	\includegraphics[width=\columnwidth]{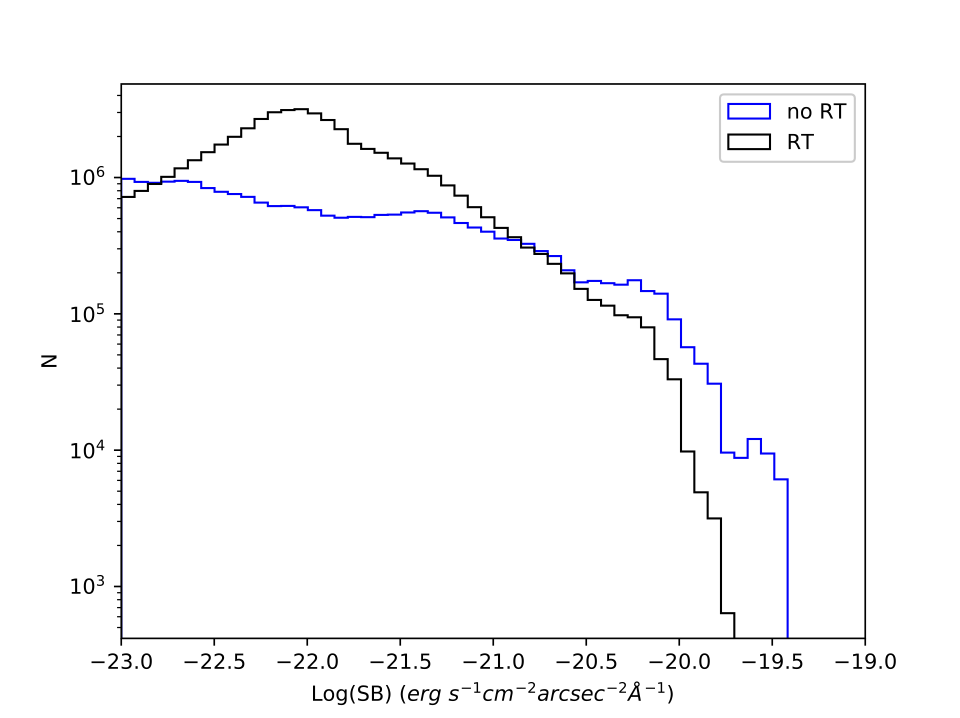}
    \caption{Impact of radiative transfer on surface brightness. Distributions of surface brightness in a box of sidelength 1250 kpc with and without radiative transfer included. Box is centered on the most massive halo in the L25n512NF simulation and has 500 pixels on a side. Radiative transfer can smear the signal so as to reduce the surface brightness in the brightest parts of the box by a factor of $\sim$2.}
    \label{fig:rtnort}
\end{figure}



%%%%%%%%%%%%%%%%%%%%%%%%%%%%%%%%%%%%%%%%%%%%%%%%%%

% Don't change these lines
\bsp	% typesetting comment
\label{lastpage}
\end{document}